\documentclass[12pt]{iopart}
\usepackage{iopams} 
\usepackage{bm}

\usepackage{graphicx}
\newcommand{\beq}{\begin{equation}}
\newcommand{\beqn}{\begin{eqnarray}}
\newcommand{\eeq}{\end{equation}}
\newcommand{\eeqn}{\end{eqnarray}}

\def\be{\begin{equation}}
\def\ee{\end{equation}}
\def\bea{\begin{eqnarray}}
\def\eea{\end{eqnarray}}
\begin{document}

\title{Non-Gaussianity of the density distribution in accelerating
universes}

\author{Takayuki Tatekawa$^{\dag,\ddag,\|}$ and Shuntaro Mizuno$^{\dag}$}

\address{\dag Department of Physics, Waseda University, 3-4-1 Okubo,
 Shinjuku, Tokyo 169-8555, JAPAN}
\address{\ddag Department of Physics, Ochanomizu University, 2-1-1 Otsuka,
Bunkyo, Tokyo 112-8610, JAPAN}
\address{$\|$ Advanced Research Institute for Science and Engineering,
Waseda University, 3-4-1 Okubo, Shinjuku,
Tokyo 169-8555, JAPAN}

\begin{abstract}
According to recent observations, the existence of the
dark energy has been considered. 
Even though we have obtained the constraint of the 
equation of the state for dark energy ($p = w \rho$)
as $-1 \le w \le -0.78$ by combining WMAP data
with other astronomical data, in order to pin down $w$,
it is necessary to use other independent observational tools.
For this purpose, we consider the $w$ dependence of the 
non-Gaussianity of the density distribution generated 
by nonlinear dynamics. To extract the non-Gaussianity,
we follow a semi-analytic approach based on Lagrangian linear
perturbation theory, which provides an accurate value
for the quasi-nonlinear
region. From our results, the difference of the non-Gaussianity between
$w = -1$ and $w= -0.5$ is about $4 \%$ while that between
$w = -1$ and $w= -0.8$ is about $0.9 \%$. For the highly non-linear 
region, we estimate the difference by combining 
this perturbative approach with N-body simulation executed for our 
previous paper. From this, we can expect the difference
to be more
enhanced in the low-$z$ region, which suggests that the non-Gaussianity
of the density distribution potentially plays an important role for extracting
the information of dark energy.
\end{abstract}


\maketitle

\section{Introduction}\label{sec:intro}

According to recent observations for type Ia
Supernovae~\cite{Perlmutter99},
the expansion of the Universe is accelerating.
Combining measurements of Cosmic Microwave Background
Radiation~\cite{WMAP03} and recent galaxy redshift survey
~\cite{Tegmark04}, researchers have concluded the Universe is almost flat, and
we are forced to recognize the existence of 
a cosmological constant, or a kind of dark energy
whose value is almost the same order of magnitude as
the present density of the Universe~\cite{WMAP}.
From the viewpoint of particle physics, however, it is quite
difficult to explain such a tiny value that is $14$ orders of
magnitude smaller than the electroweak scale. The failure
of theory to explain the present value of the 
cosmological constant is known as the
``cosmological constant problem" ~\cite{Weinberg89}.

In order to avoid this problem, many dark energy models 
have been proposed from various contexts. 
Roughly speaking, they can be classified into two approaches:
one is to modify the gravitational law and the other
is to introduce some exotic form of energy.
For familiar examples,
$1/R$ gravity models~\cite{1/R} and braneworld models~\cite{brane}
belong to the former, while quintessence models~\cite{quint,cquint},
k-essence models~\cite{kinet} and phantom models~\cite{phantom} 
belong to the latter. 

Regardless of the attempts mentioned above, it is fair to say that a
satisfactory explanation for dark energy has not yet been
obtained. On the other hand,
from the phenomenological viewpoint, 
it is important to constrain the effective equation of state
of the dark energy $w$ by observations. Even though we
have obtained the constraint as $w < -0.78$ (95 \% confidence limit
assuming $w \ge -1$)~\cite{WMAP} by combining WMAP data
with other astronomical data, in order to pin down $w$,
it is necessary to use other independent observational tools.

For this purpose, we consider the statistical properties of 
the large-scale structure of the Universe based on the
probability distribution function (PDF) of the cosmological
density fluctuations. In the standard picture, the PDF of the 
primordial density fluctuations originated from quantum fluctuations
that were stretched to large comoving scales during the
inflation phase and are assumed to be a random Gaussian.
It is, however, well known that
even though PDF remains Gaussian as long as the density 
fluctuation is in the linear regime, it significantly deviates
from the initial Gaussian shape once the non-linear stage is 
reached.

Regarding the shape of the PDF, it has been shown that in standard 
cold dark matter (SCDM) models, the PDF is fairly approximated by
lognormal distribution in a weakly non-linear regime ~\cite{Kofman94},
while the lognormal PDF does not fit well in a highly non-linear
regime~\cite{Yokoyama96}.
After that, Plionis {\it et al.}
\cite{Plionis95} and Borgani {\it et al.}~\cite{Borgani95}
 analyzed the
non-Gaussianity of the cluster distribution for several
dark matter models including the
low-density flat cold dark matter ($\Lambda$CDM) model by which
the difference of non-Gaussianity between
SCDM and $\Lambda$CDM models was clearly shown.
(The comparison among them by N-body simulations for the 
quasi-nonlinear stage was also done in~\cite{Kayo01}.)
In this paper, furthermore, we intend to investigate the
dark energy model dependence for the non-Gaussianity
of the PDF, in which we consider the simple 
constant-$w$ dark energy models with $w=-0.5, -0.8, -1.0, -1.2$.

In order to analyze the $w$-dependence of the non-Gaussianity
of the PDF, we follow a semi-analytic approach
based on Lagrangian linear perturbation theory for the evolution
of the density fluctuation.
This description provides 
relatively accurate values even in a quasi-linear regime
in the structure formation scenario
~\cite{Kofman94,Zeldovich70,Arnold82,Buchert89,Shandarin89,Paddy93A,
Paddy93B,Coles95,Sahni95,Jones04,Tatekawa05A,Paddy05,Yoshisato05}.
(Especially, for the case of the dust fluid, this approximation
is called the Zel'dovich approximation (ZA)).
Especially, in many suitable situations, 
it was shown that the Lagrangian approximation
describes the evolution of density fluctuation better than
the Eulerian approximation~\cite{Munshi94,Sahni96,Yoshisato98,
Doroshkevich73}.

Another aim of this paper is the extension of our
previous paper~\cite{Tatekawa05B} by including the dependence
of the cosmic expansion of the background.
This paper is organized as follows.
In Sec.~\ref{sec:background}, we briefly present 
the evolution equations for the background fluid
including the dark energy information. We then
also present the perturbation equations based on 
Lagrangian linear perturbation theory 
(in Sec.~\ref{sec:perturbation}) and statistical quantities
to investigate non-Gaussianity 
(in Sec.~\ref{sec:NonGauss}). 
In Sec.~\ref{sec:numerical} we provide the results and 
discuss the validity of our perturbative approach.
Section \ref{sec:summary} is devoted to conclusions.

\section{Background}
\label{sec:background}

Here we briefly introduce the evolution equations for the background
including the dark energy information.
If we assume a homogeneous and isotropic
Universe, the cosmic expansion law is described by the
Friedmann equations and the energy conservation equation.
\begin{eqnarray}
H^2 &=& \frac{8 \pi G}{3} \rho
 - \frac{\mathcal{K}}{a^2}
 \,, \label{eqn:Friedmann-00} \\
\frac{\ddot{a}}{a} &=& - \frac{4 \pi G}{3}(\rho+3p)
\,,\label{eqn:Friedmann-ii} \\
\dot{\rho} &+& 3H (\rho + p) =0
\,, \label{eqn:conserve}
\end{eqnarray}
where $a$ is a scale factor of the Universe, $H = \dot{a}/a$ is its Hubble
parameter, $\mathcal{K}$ is a curvature constant, and $p$ and $\rho$
are the total pressure and total energy of matter fields.

As for matter fields, since we are interested in a stage
much later than the
radiation-matter equality, we consider matter fluid and
dark energy, i.e., $\rho = \rho_{m} + \rho_{DE}$, where
$\rho_{m}$ and $\rho_{DE}$ are the 
matter fluid energy densities matter fluid and
dark energy, respectively. In order to obtain the evolution of
the energy densities for each component,
we must specify the properties of dark energy which are strongly dependent
on dark energy models.

Even though there are some dark energy models in which dark energy
couples explicitly to ordinal matter~\cite{cquint}, 
dark energy interacts with
ordinal matter only gravitationally in many models. We consider the case
in which the energy conservation of each component holds independently.
For these situations, since the pressure of the matter is negligible,
its energy density evolves as usual:
\begin{equation}
\rho_m \propto a^{-3} \label{eqn:rho-M} \,.
\end{equation}

As for the evolution of dark energy, we can use the fact that
it is possible to model dark energy as fluid in many dark energy models
~\cite{Weinberg89}.
For this case, the dynamical properties of dark energy
are determined through its effective equation of state $w$:
\begin{equation}
p_{DE} = w(\rho_{DE}) \rho_{DE},
\end{equation}
where, in general, $w$ is an
analytic function of the energy density, or equivalently, the
scale factor. In principle, the evolution of energy density is
determined by integrating the energy conservation equation:
\begin{equation}
\rho_{DE} (a) = \rho_{i,DE} e^{-3 \int_{a_i}^{a} d \ln a [w(a) + 1]},
\end{equation}
where $\rho_{i,DE}$ is an integration constant.
The condition for the acceleration of the cosmic expansion
is given as
\begin{equation}
w < -\frac{1}{3},
\end{equation}
where, if we are not to violate the strong energy condition, we should further
impose
\begin{equation}
-1 \le w \,.
\end{equation}

However, since it is technically difficult to specify the time dependence
of $w$ by observations in near future, we concentrate on the case for constant
$w$, in which case dark energy density evolves as
\begin{equation} \label{eqn:rho-DE}
\rho_{DE} \propto a^{-3(w+1)}  \,.
\end{equation}

Assuming flat spacetime ($\mathcal{K}=0$) and
using Eqs.~(\ref{eqn:rho-M}) and (\ref{eqn:rho-DE}),
we can rewrite the constraint Eq.~(\ref{eqn:Friedmann-00})
as
\begin{equation}
H^2 = H_0^2 \left [ \Omega_{m0} a^{-3} + \Omega_{DE0}
 a^{-3(w+1)} \right ] \,, \label{eqn:Friedmann-Omega}
\end{equation}
where we have defined the density parameters for the matter and the
dark energy:
\begin{equation}
\Omega_{(m,\;\;DE)} \equiv \frac{8 \pi G}{3 H^2} \rho_{(m,\;\;DE)}
\label{eqn:def-Omega} \,.
\end{equation}
The subscripts $0$ denote the corresponding quantities are evaluated
at present.

Under this constraint, we solve the Friedmann equation
(Eq.~(\ref{eqn:Friedmann-ii})).

\section{Perturbation}
\label{sec:perturbation}

Next, we introduce the evolution equations for the
density fluctuation of matter.
Since we are interested in a scale much smaller than
the horizon scale and non-relativistic matter fluid,
we apply a Newtonian treatment.
Notice that after this, $\rho$ denotes the mass density,
even though we used the same character as in the previous section.
Here, we start with comoving coordinates along the cosmic
expansion characterized by the solution $a(t)$ of
Eq.~(\ref{eqn:Friedmann-Omega}):
\begin{equation}
\bm{r} = a(t) \bm{x} \,,
\end{equation}
where $\bm{r}$ and $\bm{x}$ are physical coordinates and comoving
coordinates, respectively.

In the comoving coordinates,
the basic equations for cosmological fluid are described
by the continuity equation, the Euler equation and the
Poisson equation:
\begin{eqnarray}
\frac{\partial \delta}{\partial t} &+& \frac{1}{a}
\nabla_x \cdot \{ \bm{v}
(1+\delta) \} = 0 \,, \label{eqn:comoving-conti-eq} \\
\frac{\partial \bm{v}}{\partial t} &+& \frac{1}{a} (\bm{v} \cdot \nabla_x)
\bm{v} + \frac{\dot{a}}{a} \bm{v} = \frac{1}{a} \tilde{\bm{g}}
\,, \label{eqn:comoving-Euler-eq} \\
\nabla_x \cdot \tilde{\bm{g}} &=& - 4 \pi G \rho_{b} a \delta \,,
\label{eqn:comoving-Poisson-eq}
\end{eqnarray}
where $\delta \equiv \frac{\rho - \rho_{b}}{\rho_{b}}$ is the density
fluctuation, $\rho_{b}$ is background
density of the matter,
$\bm{v}$ is the peculiar velocity,
$\tilde{\bm{g}} = -\frac{1}{a} \nabla_x \Phi$
is the peculiar acceleration and $\Phi$ is the gravitational potential.
Since we note non-relativistic fluid as matter, we can ignore
the pressure and omit the suffix $m$.

In this paper, for the perturbation, we adopt
the Lagrangian picture rather than the Eulerian picture
since we can extract the quasi-nonliniear nature of
the structure formation even if we consider only the linear perturbation
\cite{Munshi94,Sahni96,Yoshisato98}.
For this purpose, it is necessary to define the comoving Lagrangian
coordinates $\bm{q}$ in terms of the comoving Eulerian coordinates
$\bm{x}$ as:
\begin{equation} \label{eqn:x=q+s}
\bm{q} = \bm{x} + \bm{s} (\bm{x},t) \,,
\end{equation}
where $\bm{s}$ is the displacement vector denoting the deviation from
homogeneous distribution.
While in Eulerian perturbation theory,
the density fluctuation $\delta$ is regarded as a perturbative quantity;
in Lagrangian perturbation theory, the displacement vector $\bm{s}$
is regarded as a perturbative quantity.
$\bm{s}$ can be decomposed
to the longitudinal and the transverse modes:
\begin{eqnarray}
\bm{s} &=& \bm{s}^L + \bm{s}^T \,, \\
\nabla \times \bm{s}^L &=& \bm{0} \,, \\
\nabla \cdot \bm{s}^T &=& 0 \,,
\end{eqnarray}
where $\nabla$ means the Lagrangian spacial derivative.

In Lagrangian coordinates, since we can solve the continuous
Eq.~(\ref{eqn:comoving-conti-eq}) from Eq.~(\ref{eqn:x=q+s}) exactly,
we can obtain the density fluctuation as

\begin{equation} \label{eqn:L-exactrho}
\delta = J^{-1}-1\,,
\end{equation}
where $J$ is the determinant of the Jacobian of the mapping
between $\bm{q}$ and $\bm{r}$: $\partial \bm{r} / \partial \bm{q}$.
It is worth noting that in Eq.~(\ref{eqn:L-exactrho}), the density
fluctuation is given in a formally exact form, even though we
keep only the linear term in this paper.

The peculiar velocity appearing in Eqs.~(\ref{eqn:comoving-conti-eq})
and (\ref{eqn:comoving-Euler-eq}) can also be expressed in terms
of the Lagrangian coordinate variable $\bm{s}$ as
\begin{equation}
\bm{v}=a \dot{\bm{s}} \label{eqn:L-velocity} \,.
\end{equation}

The basic equations we shall solve are the linearized version of
Eqs.~(\ref{eqn:comoving-conti-eq}), (\ref{eqn:comoving-Euler-eq}) and
(\ref{eqn:comoving-Poisson-eq}) in which $\delta$ and $\bm{v}$ are expressed
with $\bm{s}$ from Eqs.~(\ref{eqn:L-exactrho}) and
(\ref{eqn:L-velocity}).
The solution of the Lagrangian perturbation can be separated
into a time-dependent part and a position-dependent part as:
\begin{equation}
\bm{s}^{(1)} (t, \bm{q}) = D(t) \bm{S}^{(1)} (\bm{q}) \,.
\end{equation}
Since it is well known that there are no growing transverse mode solutions
in the linear perturbation, in this paper
we consider only the longitudinal mode in which the velocity field
is irrotational. From the Kelvin circulation theorem,
this is quite natural if it is generated only by the action of gravity.

The evolution equation for
first-order solution is written as
\begin{equation}
\ddot{D} + 2 \frac{\dot{a}}{a} \dot{D}
 - 4\pi G \rho_b D =0 \, \label{eqn:Lagrange-D},
\end{equation}
while the spacial component is given by the initial condition.
Zel'dovich derived a first-order solution of the longitudinal mode
for dust fluid~\cite{Zeldovich70} which provides the relation
between the density fluctuation and the Lagrangian displacement as
\begin{equation}
\delta(\bm{q}) = -\nabla \cdot \bm{s}^{L~(1)} \,.
\label{rel_bet_delta_s}
\end{equation}
It is worth noting that even though Eq.~(\ref{rel_bet_delta_s})
is obtained by linear perturbation, the value of $\delta$
remains accurate even in the quasi-nonlinear regime.

For the low density flat Universe with dark energy whose equation
of state is $w=-1$, i.e.,
\begin{equation}
\Omega_m + \Omega_{DE} = 1, ~~
\Omega_{DE} \equiv \frac{\Lambda}{3H^2} = {\rm const.}\,,~~
\frac{8\pi G}{3} \rho_{DE} = \Lambda\,,
\end{equation}
the growing mode first-order solution can be expressed
with the analytic function.
\begin{eqnarray}
D(t) &=& \frac{h}{2} B_{1/h^2} \left(\frac{5}{6},
 \frac{2}{3} \right ) \,, \\
h &=& \frac{H(t)}{\sqrt{\Lambda/3}} \,,
\end{eqnarray}
where $B_{1/h^2}$ is an incomplete Beta function:
\begin{equation}
B_z (\mu, \nu) \equiv \int_0^z p^{\mu-1} (1-p)^{\nu-1} {\rm d} p \,.
\end{equation}

For other dark energy models, we shall solve Eq.~(\ref{eqn:Lagrange-D})
numerically.

In order to avoid the divergence of the density fluctuation in the limit of
large $k$, however, just for a technical reason,
it is necessary to consider the density field $\rho(\bm{x};R)$
at the position $\bm{x}$ smoothed over the scale $R$,
which is related to the unsmoothed density field
$\rho(\bm{x})$ as

\begin{eqnarray}
\rho(\bm{x};R) &=& \int d^3 \bm{y} W(|\bm{x}-\bm{y}|;R) \rho(\bm{y})
\nonumber\\
&=& \int \frac{d^3 \bm{k}}{(2\pi)^3} \tilde{W} (kR)
\tilde {\rho} (\bm{k}) e^{-i \bm{k}\cdot \bm{x}} \,,
\end{eqnarray}
where $W$ denotes the window function and
$\tilde{W}$ and $\tilde{\rho}$ represent the Fourier transforms
of the corresponding quantities.
In this paper, we adopt the top-hat window function,
\begin{eqnarray}
\tilde{W} = \frac{3(\sin x - x \cos x)}{x^3}\,.
\end{eqnarray}

Then, the density fluctuation $\delta(\bm{x};R)$
at the position $\bm{x}$
smoothed over the scale $R$ can be constructed
in terms of $\rho(\bm{x};R)$. For simplicity we use
$\delta$ to denote $\delta(\bm{x};R)$
unless otherwise stated.

\section{Non-Gaussianity of the density fluctuation}
\label{sec:NonGauss}

In order to analyze the statistics, we introduce a one-point probability
distribution of the density fluctuation field $P(\delta)$
(PDF of the density perturbation) which denotes the probability of
obtaining the value $\delta$.
If $\delta$ is a random Gaussian field, the PDF of the density perturbation
is determined as
\begin{eqnarray*}
P(\delta) = \frac{1}{(2\pi \sigma^2)^{1/2}} e^{-\delta^2/2\sigma^2},
\end{eqnarray*}
where $\sigma \equiv \left < \left (\delta-
 \left <\delta \right > \right )^2
 \right > $ and $\left <\;\;\right >$ denotes the spacial average.

If the PDF deviates from Gaussian distribution, in order to specify
it, it is necessary to introduce the following higher-order statistical
quantities~\cite{Kofman94,Peebles80,Peacock}:
\begin{eqnarray*}
\mbox{skewness} &:& \gamma = \left < \left (\frac{\delta-
 \left <\delta \right >}{\sigma}
 \right )^3 \right > \,, \\
\mbox{kurtosis} &:& \eta = \left < \left (\frac{\delta-
 \left <\delta \right >}{\sigma}
 \right )^4 \right > - 3 \,,
\end{eqnarray*}
which mean the display asymmetry and non-Gaussian degree of
``peakiness,'' respectively.

\begin{figure}[tb]
\centerline{
\includegraphics[height=7cm]{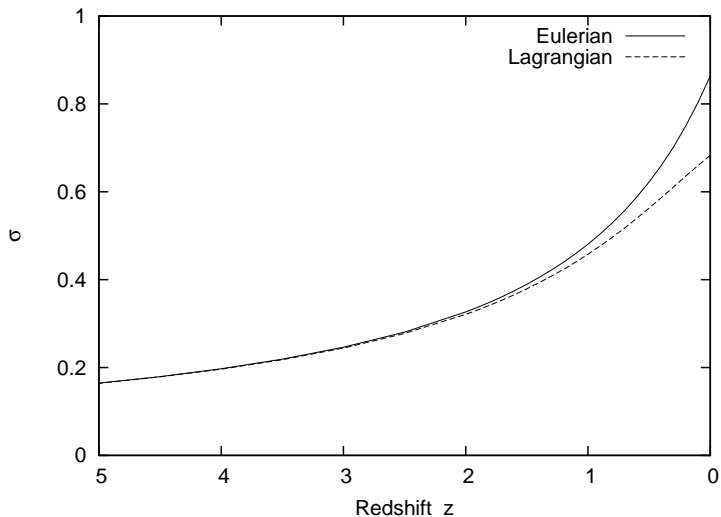}
}
\caption{The dispersion of the density fluctuation
in both Eulerian and Lagrangian linear perturbation
 for the model with $w = -1$. Here we choose
the smoothing scale $R=8 h^{-1} \mbox{Mpc}$
for the top-hat window function.
After $z \sim 5$, the difference of the dispersion
  appears between Eulerian and Lagrangian
 in which Eulerian linear perturbation no longer provides 
 accurate values.
Therefore we can expect to obtain the nontrivial
information for the skewness and the kurtosis
in this quasi-nonlinear region
from Lagrangian perturbation
 as long as it provides accurate values.}
\label{fig:Dust-sigma-EL}
\end{figure}

In Eulerian perturbation theory, the Gaussianity
of the PDF completely conserves if we start with Gaussian distribution,
because the density fluctuation is described by the product of
the time and the spacial component and
the spacial component never changes as long as we keep only linear terms.
Once nonlinear terms are considered, however, its PDF deviates from the
initial Gaussian shape because of the strong nonlinear mode coupling
and the nonlocality of the gravitational dynamics.
On the other hand, since in Lagrangian perturbation theory,
the quasi-nonlinear information in the sense of the Eulerian picture
can be extracted even by linear perturbation, we can expect
to obtain nontrivial information for the skewness and the kurtosis
after $z \sim 5$. See Fig.~\ref{fig:Dust-sigma-EL}.

\section{Numerical results}
\label{sec:numerical}

In this section, after mentioning the conditions we impose, we present
our results.

We consider here the Gaussian density field generated by COSMICS~\cite{COSMICS},
and
we set the value of the density fluctuation and the peculiar velocity
to be those obtained by Lagrangian linear perturbation with the
time evolution of the background to be $a \propto t^{2/3}$.
This is reasonable because at this time,
the dark energy did not dominate the CDM.
Accordingly, both the skewness and the kurtosis
are less than $10^{-2}$ at initial ($z \sim 30$).

For the computation of the Lagrangian perturbation, we set
the parameters as follows:
\begin{eqnarray*}
\mbox{Number of grids} &:& N=128^3 \,, \\
\mbox{Box size} &:& L=128 h^{-1} \mbox{Mpc}
 ~~(\mbox{at}~a=1)  \,.
\end{eqnarray*}
Then we impose a periodic boundary condition.

The cosmological parameters
at the present time ($a=1$) are given
by~\cite{WMAP}
\begin{eqnarray}
\Omega_m &=& 0.27 \,, \\
\Omega_{DE} &=& 0.73 \,, \\
H_0 &=& 71~ \mbox{[km/s/Mpc]} \,, \\
\sigma_8 &=& 0.84 \,.
\end{eqnarray}

For the dark energy models,
we set several equations of state.
\begin{equation}
p= w \rho, ~~ (w=-0.5, -0.8, -1, -1.2) \,,
\end{equation}
where $w=-1$ corresponds to the cosmological constant. Even though
$w=-0.5$ has already been excluded, we calculate this just for 
comparison. In the case of $w=-1.2$, as mentioned before, the strong
energy condition is violated.

As for the smoothing scale $R$ for the top-hat window function,
we set $8 h^{-1} \mbox{Mpc}$  in the comoving Eulerian coordinates
at the present time ($a=1$). Because we apply Lagrangian
linear perturbation, we must consider the validity of
the perturbation. As we will show later, it is reasonable
that we set the smoothing scale $R=8 h^{-1} \mbox{Mpc}$
if we are 
to discuss the structure formation with the perturbation.

Based on $\delta$ calculated from these conditions,
we compute the statistical quantities, i.e., the dispersion, the skewness
and the kurtosis
at 51 time slices from $z=5$ to $z=0$ whose time intervals
are given as $\Delta z=0.1$.

Before presenting the results, we must remind ourselves that
since these calculations are based on Lagrangian linear perturbation
theory, we can trust the results until caustics are produced,
i.e.,
when $\sigma$ becomes $\sim 1$.

\begin{figure}[tb]
\centerline{
\includegraphics[height=7cm]{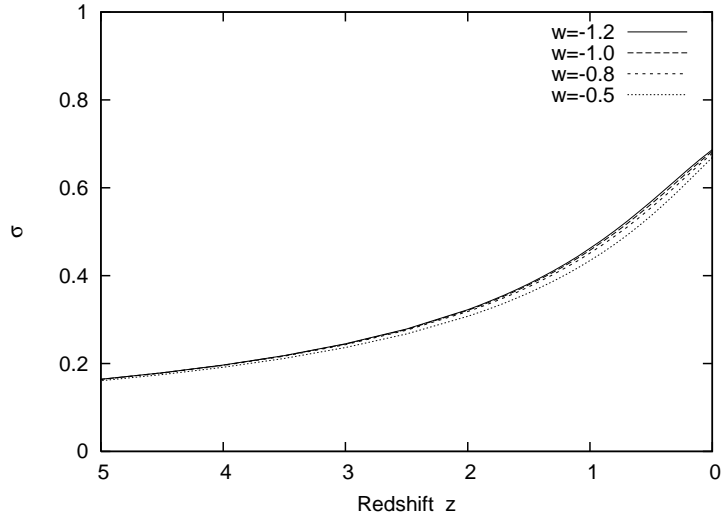}
}
\caption{The dispersion of the density fluctuation
in several dark energy models.
The growth of the dispersion 
continues monotonously.
The difference of the dispersion between $w=-1$ and
$w=-0.8$ is about $0.9 \%$ at $z=2$ until which we regard 
 Lagrangian linear perturbation as valid. In the same way,
the difference of the dispersion between $w=-1$ and
$w=-1.2$ is about $0.4 \%$ at $z=2$.
}
\label{fig:DE-sigma}
\end{figure}

Figure~\ref{fig:DE-sigma} shows
the time evolution of the dispersion of
the density fluctuation for several
equations of state.
The growth of the dispersion monotonously continues
until $z=0$. Because we choose a relatively large value
for the smoothing length, the Lagrangian linear perturbation
still has physical meaning. 
The tendency of structure formation can be discussed
even if structure formation can no longer be fully
described with linear approximation.

We can identify the difference of the
dispersion among several values of $w$. As the value of
$w$ becomes larger (approaches to 0), the dispersion becomes smaller.
This can be explained as follows:
For the model with larger $w$, dark energy
dominates at earlier era, and the expansion of
the Universe starts to accelerate at earlier era,
while it can be shown that the growth of the
density fluctuation is smaller in the accelerating stage
than in the matter-dominant stage.

Here we consider high-$z$ region, i.e., $z \ge 2$. 
In our previous paper~\cite{Tatekawa05B},
we compared the evolution of the dispersion between
N-body simulation and Lagrangian linear perturbation.
There, 
the difference of the dispersion, the skewness, and
the kurtosis between N-body simulation
and the Lagrangian linear perturbation
stayed small until $\sigma \simeq 0.3$.
From the past analyses, the Lagrangian linear perturbation
describes structure formation rather well until $z \ge 2$,
where the dispersion is $\sigma \simeq 0.3$.

\begin{figure}[tb]
\centerline{
\includegraphics[height=7cm]{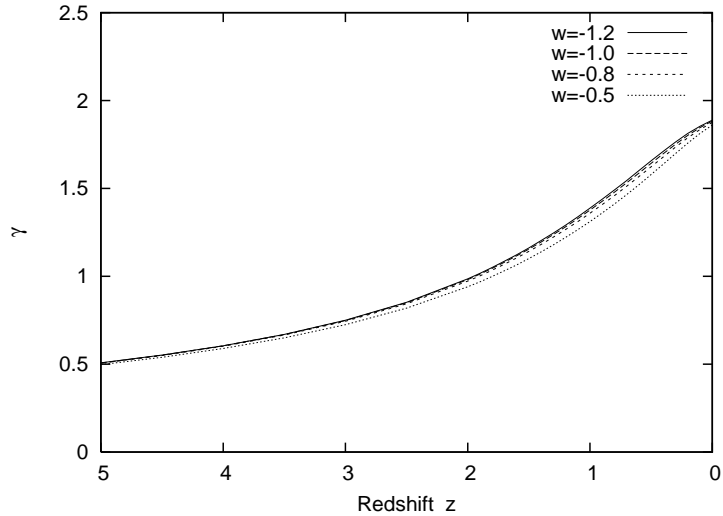}
}
\caption{The skewness of the density fluctuation
in several dark energy models. 
The growth of the skewness continues monotonously.
Although we can easily distinguish the case where $w=-0.5$ from
the results,
the difference of the skewness between $w=-1$ and
$w=-0.8$ is about $0.9 \%$ at $z=2$ until which we regard  
 Lagrangian linear perturbation as valid. In the same way,
the difference of the skewness between $w=-1$ and
$w=-1.2$ is about $0.4 \%$ at $z=2$.
}
\label{fig:DE-skew}
\end{figure}

\begin{figure}[tb]
\centerline{
\includegraphics[height=7cm]{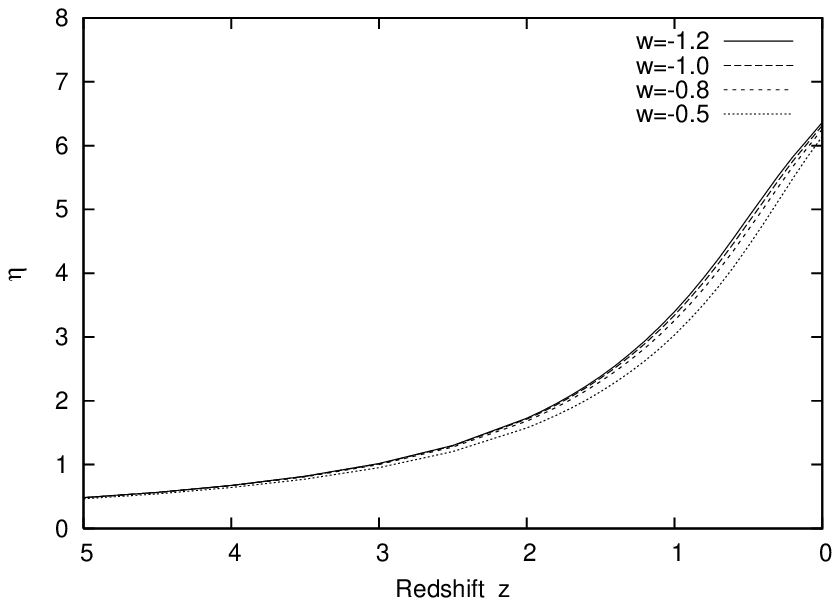}
}
\caption{The kurtosis of the density fluctuation
in several dark energy models.
The growth of the kurtosis monotonously continues.
As well as the distinction of the skewness,
the difference of the kurtosis between $w=-1$ and
$w=-0.8$ is about $1.8 \%$ at $z=2$ until which we regard the 
Lagrangian linear perturbation is valid. In the same way,
the difference of the kurtosis between $w=-1$ and
$w=-1.2$ is about $0.9 \%$ at $z=2$.
}
\label{fig:DE-kurt}
\end{figure}

Withrespect to non-Gaussianity, 
Figs.~\ref{fig:DE-skew} and \ref{fig:DE-kurt}
show the evolution of the skewness and the kurtosis,
respectively.
Like the dispersion, the growth of the skewness
and the kurtosis continues monotonously.

For the reason mentioned above,
our computations seem to be reliable until $z \simeq 2$.
For the case where $w=-0.5$, the value of both
the skewness and the kurtosis is obviously less than
that for the cases where $w=-1.2$, $w=-1.0$ and $w=-0.8$
(about $4 \%$ and about $8 \%$, respectively).
On the other hand, the differences of both the skewness and
the kurtosis among $w = -1.2$, $w=-1$ and $w=-0.8$ are
less than $2 \%$ at $z=2$. Even though these are very small values,
they are of almost the same order as the relative ratio
of the dispersions. Therefore it can be expected that
 non-Gaussianity can play a potentially important role
in distinguishing the value of $w$ from the observation of
 high-z galaxies.
At that stage, the fitting functions for $\gamma_w (z)$,
$\eta_w (z)$ for the quasi-nonlinear stage are useful.
Since it seems that the growth of the dispersion and
the non-Gaussianity are related, it is worth trying
to find the functions based on our results.

Finally, if we want to know precise information
for non-Gaussianity after $z \sim 2$, we must consider
nonlinear dynamics by such as N-body simulations.
Even though we do not compute them in this paper, judging from
Fig.1 in our previous paper, \cite{Tatekawa05B},
in which we compare the results of 
Lagrangian perturbation and N-body simulation in
the Einstein-de Sitter
Universe, we can expect the actual non-Gaussianity is strengthened
by a order of magnitude. If this happens also for accelerating Universes,
it becomes easier to distinguish $w$ from the observations of
the near galaxies, which is worth investigating.

\section{Summary} \label{sec:summary}

To clarify the nature of the dark energy is, without doubt,
one of the most important tasks in modern cosmology.
From the phenomenological viewpoint, 
it is important to constrain the effective equation of state
of dark energy $w$ by observations. Even though we
have obtained the constraint as $w < -0.78$ (95 \% confidence limit
assuming $w \ge -1$)~\cite{WMAP} by combining WMAP data
with other astronomical data, to pin down $w$,
it is necessary to use other independent observational tools.

For this purpose, we consider the non-Gaussianity of
the density fluctuation which is generated by the non-linear
dynamics. In this paper, we follow a semi-analytic approach
based on Lagrangian linear perturbation theory for the evolution
of the density fluctuation.
In this theory, we can extract the quasi-nonlinear feature
of the density fluctuation even though we keep only linear terms.
In terms of the density fluctuation we obtain,
we compute the skewness and the kurtosis which are 
statistical quantities denoting the degree of non-Gaussianity
as well as the dispersion, which is the only parameter
for Gaussian distribution.
By considering several constant-$w$ dark energy models,
we present the $w$ dependence of non-Gaussianity.
Because of the validity of the perturbative approach,
we regard the results until $z \simeq 2$ as reliable.

According to our calculation, at fixed time $(z)$, the dispersion becomes
smaller as $w$ becomes larger (approaches to 0).
This is because the larger $w$ is, the earlier the time
at which the expansion of the Universe starts to accelerate and
the growth of the density fluctuation
is smaller than in the matter-dominant Universe.

The skewness and the kurtosis also become smaller as
$w$ becomes larger (approaches to 0), which suggests
they have some correlation with the dispersion.
The difference of non-Gaussianity between $w=-1$ and $w=-0.5$
is obvious (about $4 \%$) for $2<z<5$, while
it is hard for $w$ to distinguish $-1.2$, $-1$ or $-0.8$.
The differences of both the skewness and the kurtosis
between $w=-1$ and $w=-0.8$ as well as
$w=-1.2$ and $w=-1.0$ are quite small (about $0.9 \%$),
while the correspondent differences of the
dispersion are the almost same as these.
To specify the value of $w$,
the fitting functions for $\gamma_w (z)$,
$\eta_w (z)$ for the quasi-nonlinear stage
which can be constructed from our results are useful.

For regions nearer than $z=2$, it is necessary to pick up
non-linear information for obtaining the degree of 
non-Gaussianity. Even though we have not computed these in this
paper, from our previous results comparing N-body simulation
with Lagrangian perturbation theory, it can be expected that
the actual value of non-Gaussianity becomes larger than
the value obtained by Lagrangian linear perturbation.
For this case, it becomes easier to distinguish the value of
$w$ from the observation of near galaxies, which is worth
continuing to consider.

It is necessary to mention the possibility
of examining our results.
Recently, several galaxy redshift survey projects have been
progressing~\cite{Hawkins03,Abazajian04,Gerke04}. In these projects,
many galaxies within a region ($z<0.3$ for 2dF,
$z<0.5$ for SDSS, and $0.7 < z < 1.4$ for DEEP2) have
been observed. These projects show the latest distribution of
galaxies, which form strongly nonlinear structure.
For the next generation of spectroscopic surveys,
many high-$z$ galaxies will be observed. For example, the
Cosmic Inflation Probe (CIP) project~\cite{CIP} is planned
to detect objects between $3 < z < 6.5$. Another project,
the Kilo-Aperture Optical Spectrograph (KAOS),  has also been
proposed~\cite{KAOS}. One of the primary scientific purposes
of KAOS is the determination of the equation of state
of dark energy by objects until $z < 3$. 
Another purpose of KAOS is to observe the growth of structure.
The projects leaders propose that
the linear growth factor of matter density fluctuations
can also provide sensitivity to dark energy properties.

The skewness and the kurtosis of the distribution
of galaxies in the SDSS Early Data Release had been computed
~\cite{SDSS-NonGauss}.
The results suggested the SDSS imaging data
can enable us to determine the skewness and the kurtosis up
to 1\% and less than 10\%. Therefore in future projects, 
we can expect to determine the skewness and the kurtosis
with high accuracy.
From our results, the kurtosis of the density fluctuation
seems more sensitive for $w$ than the dispersion of
the fluctuation. In other words, although the dispersion
reflects the linear growth factor, the kurtosis
shows the nature of nonlinear growth in addition to
the linear growth. Therefore, structual non-Gaussianity of
would show more detailed information
on  the nature of dark energy.
It is worth noting that for such a high-$z$ region, the results
obtained in this paper based on our perturbative approach 
will play an important role.

 Finally, it is worth noting the primordial
non-Gaussianity which is generated by inflation.
From WMAP observations, a limit for
the non-linear coupling parameter has been established
\cite{WMAP-NonGauss} which 
does not deny the existence of non-Gaussianity
in the primordial density fluctuation.
Even though we cannot disentangle them generally,
we can calculate and compare  non-Gaussianity
of large-scale structure from both Gaussian
and non-Gaussian initial conditions.
This requires further investigation, and we hope to report
results in a separate publication.

\ack

We thank to Masahiro Morikawa for useful discussion.
TT was supported by the Grant-in-Aid for Scientific
Research Fund of the Ministry of Education, Culture, Sports, Science
and Technology, Japan (Young Scientists (B) 16740152).
SM was supported by the
Grant-in-Aid for Scientific
Research Fund of the Ministry of Education, Culture, Sports, Science
and Technology, Japan (Young Scientists (B) 17740154).

\end{document}